# A model-based approach for transforming InSAR-derived vertical land motion from a local to a global reference frame

Mahmoud Reshadati, Manoochehr Shirzaei

*Abstract*— **Vertical land motion (VLM) observations obtained from Interferometric Synthetic Aperture Radar (InSAR) have transformed our understanding of crustal deformation processes over the past 3 decades. However, these observations are often related to a local reference frame, posing challenges for studies that require large-scale observations within a global reference frame, such as assessments of relative sea level rise and associated hazards. Here, we present a novel approach that enables transforming InSAR-derived VLM at any location worldwide to a global (e.g., International Terrestrial Reference Frame) reference frame without a direct need for GNSS (Global Navigation Satellite System) measurements. To this end, we employ a coarse resolution model of global VLM obtained by interpolating rates of all available GNSS stations over the global land areas. Our rationale is that the high-resolution InSAR-derived VLM data do not capture the long-wavelength signals present in the global VLM model. Therefore, we employ a set of 2D polynomial models to evaluate the difference between InSAR-derived VLM and the global model and then add it back to the InSAR-derived VLM. We examined the validity of our rationale using normalized power spectrum analysis and tested the effect of polynomial order on the accuracy of transformed VLM and the overall success of our approach using two datasets from Los Angeles and New York City. This approach improves the usability of InSAR-derived VLM in geophysical applications, including monitoring regional land subsidence.**

Mahmoud Reshadati's contribution is supported by the Department of Energy. Manoochehr Shirzaei thanks the support from the Department of Defense. (*Corresponding author: Mahmoud Reshadati.*)

Mahmoud Reshadati is with Department of Geosciences, Virginia Tech, Virginia Tech, Blacksburg, VA 24061, USA (e-mail: mahmoudreshadati@vt.edu ).

Manoochehr Shirzaei is with Department of Geosciences, Virginia Tech, Virginia Tech, Blacksburg, VA 24061, USA and also with United Nations University Institute for Water, Environment and Health, Richmond Hill, ON, CA (e-mail: shirzaei@vt.edu )

## I. INTRODUCTION

INTERFEROMETRIC Synthetic Aperture Radar (InSAR) provides high-resolution ground motion observations, finding extensive use in topographic mapping, surface deformation monitoring, and change detection [1], [2], [3], [4]. However, the InSAR measurements rely on an arbitrary choice of reference point (RP), a pixel with known or assumed zero displacements within the image frame, and thus, InSAR images inherently convey relative information. Assuming a zero displacement for the reference point simplifies the problem but carries inherent drawbacks. The premise of zero motion rate suggests stability, yet empirical evidence from various research [5], [6], [7], [8] indicates that such an assumption may not be valid. Even seemingly stable regions may experience subtle movements or deformations, for instance, due to rigid plate motions, that could impact the interpretation of results. Furthermore, every pixel within the InSAR image, including the reference pixel, is subject to some noise, which can propagate and affect the overall quality of the deformation map [9], [10], and the displacement time series of the other pixels [11]. Consequently, an InSAR deformation map is inherently sensitive to uncertainties associated with the RP pixels and their temporal stabilities.

To address these challenges, it is necessary to transform the InSAR measurements to a global reference frame, such as the International Terrestrial Reference Frame (ITRF) [12], [13]. This step is crucial for regional studies of vertical land motion, such as assessment of relative sea level rise that combines measurements of coastal land elevation change with sea level rise observations [14], [15], [16], [17]. To transform InSAR measurements into a global reference frame, they are often combined with that of the Global Navigation Satellite System (GNSS) [18], [19], [20]. However, GNSS networks are sparse, mainly outside the US and Europe, and thus, there may not be enough stations within the InSAR frame to perform the transformation.

Here, we present an approach to transform InSAR-derived vertical land motion (VLM) from a local to a global reference frame using a coarse resolution model of VLM generated by interpolating GNSS observations over the global lands following [21] (Fig. 1a). The authors used a processed GNSS database aligned with the ITRF to provide comprehensive maps of VLM globally. This map comprises a long-wavelength component of the surface deformation, which mainly resembles the deformation due to Glacial Isostatic Adjustment [22] as shown in Fig. 1b. We utilized the ICE-6G_C (VM5a) model developed by Peltier et al. [23] and further assessed in [22] for the GIA-induced vertical motion to address the large-scale VLM in both cities. GIA is crucial in understanding and modeling land movement because it accounts for significant changes in solid Earth deformation over tens of thousands of years, representing the VLM In regional contexts [24].

We rationale that the InSAR-derived VLM map lacks long spatial wavelength deformation components primarily due to rigid translation, rotation, and scaling of the local reference frame concerning the global one and that these long wavelengths can be evaluated and restored given an existing model of VLM at a coarser resolution such as the one provided by [21].

Therefore, we apply a least squares approach to fit a smooth polynomial of orders of degree one, two, and three, named as D1, D2, D3, respectively, to the difference between InSAR-derived VLM and the global VLM model and add this polynomial back to the InSAR-derived VLM to transform it from the local to the global reference frame. To demonstrate our approach, we used InSAR-derived VLM datasets from Los Angeles (LA) and New York City (NYC) published in the literature [25], [26]. We employ a spectral analysis to assess the success of our approach in restoring the long wavelength signals and use available GNSS datasets to evaluate the accuracy of the transformed VLM. This work advances InSAR data processing methodologies by offering a systematic approach to converting relative VLM measurements into absolute values using a coarse-resolution model of global VLM, which is suitable for large-scale studies of land subsidence.

## II. MATERIALS AND METHOD

### A. InSAR datasets

This study uses synthetic aperture radar (SAR) datasets from Sentinel-1 satellites over two major New York City and Los Angeles urban areas. The New York City dataset spans from March 2015 to August 2022 and includes 183 SAR images [26]. The Los Angeles dataset covers from February 2016 to August 2022, comprising 247 SAR images [25]. Both datasets are in descending orbits. We obtained the published line-of-sight (LOS) velocities from the literature above and converted them to vertical by dividing the LOS with the cosine of the local incidence angles, assuming horizontal motions can be modeled and removed using a plane. Both datasets are processed using the wavelet-based InSAR algorithm [25], [27], [28], [29], which implements a suite of filters to generate accurate maps of LOS velocity.

### B. Transformation framework

The framework for transforming InSAR-derived VLM rates from a local to a global frame using a coarse-resolution global VLM model is presented in Fig. 2. Given an InSAR-derived VLM rate map in a local reference frame, we oversample the global VLM rate model at the locations of the InSAR pixels. Then, we compute the difference between the InSAR VLM and the interpolated model VLM. This difference accounts for the discrepancies such as offset, rotation, and scaling inherent in the local measurements compared to the global model. Next, we use a low-order polynomial to model the VLM difference array. The general formula for a 2D polynomial is [30]:

$$\delta_i = \beta_0 + \sum_{j=1}^{k} \sum_{p=0}^{j} \beta_{j-p,p} x_{1,i}^{j-p} x_{2,i}^{p} + \varepsilon_i \quad (1)$$

For each pixel (i), $\delta_i$ represents the VLM difference between local measurements and global ones, $x_{1,i}$ and $x_{2,i}$ are the coordinates, $\beta_{j-p,p}$ is the coefficients reflecting the combined effects of $x_1$ and $x_2$, and $\varepsilon_i$ is the error term for the i-th observation point. Rewriting equation 1 in matrix formulation:

$$\Delta = X\beta + \epsilon \quad (2)$$

Which can be written as:

$$\begin{bmatrix} \delta_1 \\ \delta_2 \\ \delta_3 \\ \vdots \\ \delta_n \end{bmatrix} = \begin{bmatrix} 1 & x_{11} & x_{21} & x_{11}^2 & x_{11}x_{21} & \dots & x_{11}^k & x_{11}^{(k-1)}x_{21} & \dots & x_{21}^k \\ 1 & x_{12} & x_{22} & x_{12}^2 & x_{12}x_{22} & \dots & x_{12}^k & x_{12}^{(k-1)}x_{22} & \dots & x_{22}^k \\ 1 & x_{13} & x_{23} & x_{13}^2 & x_{13}x_{23} & \dots & x_{13}^k & x_{13}^{(k-1)}x_{23} & \dots & x_{23}^k \\ \vdots & \vdots & \vdots & \vdots & \vdots & \ddots & \vdots & \vdots & \ddots & \vdots \\ 1 & x_{1n} & x_{2n} & x_{1n}^2 & x_{1n}x_{2n} & \dots & x_{1n}^k & x_{1n}^{(k-1)}x_{2n} & \dots & x_{2n}^k \end{bmatrix} \begin{bmatrix} \beta_0 \\ \beta_{(1,0)} \\ \beta_{(0,1)} \\ \beta_{(2,0)} \\ \beta_{(1,1)} \\ \beta_{(0,2)} \\ \vdots \\ \beta_{(k,0)} \\ \beta_{(k-1,1)} \\ \vdots \\ \beta_{(0,k)} \end{bmatrix} + \begin{bmatrix} \varepsilon_1 \\ \varepsilon_2 \\ \varepsilon_3 \\ \vdots \\ \varepsilon_n \end{bmatrix} \quad (3)$$

The matrix $X$ contains coordinates, $\Delta$ is the VLM difference values, and the vector $\beta$ contains the model coefficients. We employ the least squares regression method to determine the coefficients array $\beta$:

$$\beta = (X^\top X)^{-1} X^\top \Delta \quad (4)$$

Next, we calculate the model $X\beta$, representing the modeled VLM differences at each pixel, and add it back to the InSAR-derived VLM to obtain the transformed VLM, as follows:

$$VLM^{Transformed} = VLM^{local} + X\beta \quad (5)$$

Our approach aims to obtain a VLM dataset that embodies both the regional (long-wavelength) characteristics from the coarse-resolution global VLM model and the local (short-wavelength) attributes from local InSAR-derived VLM. Note that the calculations become ill-conditioned for polynomials of higher degrees. Therefore, we only consider the polynomial degrees 1, 2, and 3.

## III. RESULTS

To evaluate our approach, we consider New York City (NYC) and Los Angeles (LA) InSAR-derived VLM as case studies (Fig. 3 and 4). We validate the performance of our approach by comparing the transformed VLM data with GNSS measurements and employing a spectral analysis. We obtained GNSS datasets in IGS14 reference frame from the Nevada Geodetic Laboratory with observation periods that overlap with the InSAR period for at least three years and cover a similar period. We define a radius of 100 meters around each GNSS station and compute the mean value of the InSAR-derived VLM for pixels inside the circle as the corresponding InSAR Value. We use several metrics, such as mean absolute error (MAE) and root mean square error (RMSE) to evaluate model performance, and Information Criteria metrics (AIC and BIC) to find the model that has the most favorable balance between model complexity and fitness (Table 1).

### A. New York City

NYC's VLM map is characterized by uplift patterns of up to 2 mm/yr and localized subsidence rates reaching up to -2 mm/yr (Fig. 3a), attributed to anthropogenic activities such as groundwater extraction and recharge and sediment compaction [26], [31], [32], [33], [34] The inset shows a bivariate plot that compares InSAR-based VLM with GNSS vertical measurements. As can be seen, the difference in reference frames causes an offset between the InSAR and GNSS measurements. The map of the global VLM model oversampled at the location of InSAR pixels is shown in Fog. 3b, which is dominated by subsidence rates up to -1 mm/yr. Figure 3c also presents the GIA effect, which includes long-wavelength subsidence signals up to -2mm/yr.

The transformed VLM into the global reference frame is shown in Fig. 4. To this end, we used three different polynomial regressions and found that the 1st-degree polynomial regression yields the lowest RMSE and MAE values, with the MAE notably reduced from 0.22 to 0.05 mm/yr. This reduction in MAE, along with a 0.07 mm/yr for the standard deviation of the difference between InSAR VLM and GNSS (Table 1), indicates the success of our transformation using $1^{st}$-order polynomial. In addition, having the lowest BIC and AIC indicates the best tradeoff between fitness and complexity of the 1st-degree model in comparison with the other models. The Bivariate plots also show a good agreement between transformed VLM and GNSS observation. Compared with the bivariate plots in Fig. 3a, we note that the offset between InSAR and GNSS is entirely removed.

Fig. 5d illustrates the data sets' empirical cumulative distribution function (CDF) curves. As seen, the CDF curve for local VLM is offset from the remaining CDFs, while after the transformation, all CDF curves shift to align with the global VLM curve. The amount of this shift (~1.8mm/yr) is consistent with the average GIA-induced VLM (~-1.7 mm/yr) in the area.

### B. Los Angeles

The VLM map of Los Angeles (Fig. 5a) shows a patchy distribution of subsidence and uplift up to 5mm/yr, attributed to tectonic and anthropogenic activities such as groundwater depletion and hydrocarbon exploration [19], [25], [35], [36], [37], [38]. The inset compares InSAR-based VLM with GNSS data, indicating a good agreement between the two methods. The map of global VLM oversampled at the location of InSAR pixels (Fig. 5b) shows minor movements of up to 0.5 mm/yr, implying small regional vertical land motion in the area. Further, figure 5c shows the GIA effect in the region with subsidence up to 0.8 mm/yr.

Among the transformed VLM maps of Los Angeles (Fig. 6), the 1st-degree polynomial regression method shows the best results by offering the lowest RMSE, MAE values, and the lowest BIC and AIC. The Bivariate Plots in Fig. 6 also show a good agreement between InSAR and GNSS observations. The CDF curves for the InSAR-derived VLM and transformed VLM align well with the CDF curve from the GNSS-derived VLM. The shift in the CDF chart after transformation is considerably smaller (~-0.4 mm/yr) than that of New York City, comparable to the average GIA rate in the region.

## IV. VLM SPECTRAL ANALYSIS

To further evaluate our approach's success, we examine the spectral contents of VLM maps, using the Fourier Transform to identify dominant spatial wavelengths [39], [40]. The rationale is that a successful transformation fuses spectral contents of local VLM with that of the global model. As seen in Fig. 7a, for the case of New York City, the global VLM model comprises significant power at long wavelengths (>10 km), which rapidly diminishes, indicating the influence of large-scale processes such as GIA. The local InSAR VLM also has high power at long wavelengths. However, towards longer wavelengths, the more negative slope of its power spectrum curve (inset in Fig. 7a), compared with that of the global model, suggests that the local InSAR VLM lacks some deformation components of the longer wavelength. The power spectrum of the transformed VLM to the global reference frame yields a balance between that of local InSAR and global model while entirely capturing the short wavelength of the local InSAR VLM. Thus, we conclude that the transformation successfully fuses the long wavelength attributes into the local VLM data while keeping the short wavelength attributes.

For Los Angeles, the global VLM model shows no significant power at long wavelengths, indicating the absence of dominant large-scale processes (Fig. 7b). In addition, the local InSAR-derived VLM and the transformed models exhibit nearly identical power spectra, suggesting that the transformation does not significantly alter the InSAR-derived VLM map. We conclude that the absence of long-wavelength components explains why the transformation has a limited impact. By comparing these two case studies, we find that when the global VLM model comprises the regional characteristics, our method effectively incorporates this long-wavelength information into the InSAR-derived VLM while preserving short-wavelength information. However, this also highlights a shortcoming of our approach. The success of our approach critically depends on how well the long wavelength signals are captured by the global VLM model, which is a function of the density of the GNSS network and interpolation methods used for creating this coarse resolution model. Thus, although our method is applicable globally, the accuracy of results, particularly outside the US and Europe, varies and needs to be accounted for when interpreting the results.

## V. Conclusions

We presented a novel approach for transforming InSAR-derived VLM from a local to a global reference frame using a pre-existing coarse-resolution model of global VLM. This method utilizes a polynomial regression model to estimate missing long-wavelength deformation signals and restore them to the local VLM. We examined two InSAR-derived VLM data sets from New York and Los Angeles to demonstrate our approach and validate the results against available GNSS observations. Where there are considerable regional-scale land motions (e.g., New York case study), our framework can adequately fuse it into the InSAR-derived VLM, obtaining a map consistent with GNSS measurements within a global reference frame. On the other hand, where there are no major long-wavelength deformation signals (e.g., Los Angeles case study), our approach does not distort the local deformation signals. We emphasize that the quality of our results critically depends on the accuracy of the coarse global VLM model. However, the presented framework allows for transforming InSAR-derived VLM to a unified global reference frame wherever they are measured, suitable for urban planning, infrastructure development, and natural hazard management.

## Data Availability

The InSAR-derived VLM maps used in this study were obtained from [25], [26]. The GNSS measurements used to validate our InSAR-based VLM results were downloaded from the Nevada Geodetic Laboratory at http://geodesy.unr.edu as MIDAS Velocity Fields in IGS14 reference frame [41]

The Global VLM model is based on [21] and the GIA model used in this study is sourced from the [23] model, available at the Permanent Service for Mean Sea Level (PSMSL) website https://psmsl.org/train_and_info/geo_signals/gia/peltier/index.php

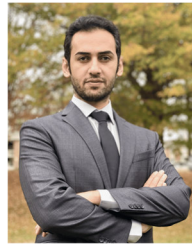

**Mahmoud Reshadati** received His B.Sc. degree in Physics from Sharif university of Technology, Tehran Iran and the M.Sc. degree in Geophysics from the university of Tehran. He is currently working toward the Ph.D. degree in the Department of Geosciences at Virginia Tech.

At this stage of academic career, his research focus is on remote sensing, signal processing, and application of data science in geophysics and geo-energy

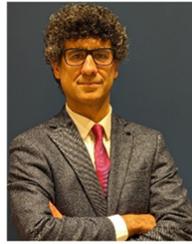

**Manoochehr Shirzaei** received an M.Sc. degree in civil/surveying engineering (Geodesy) from Tehran University, Tehran, Iran, in 2003 and a Ph.D. degree in geophysics and remote sensing from the University of Potsdam, Potsdam, Germany, in 2010. From 2007 to 2011, he was a Researcher with the German Research Center for Geosciences, Potsdam, Germany, where he developed several advanced multitemporal InSAR algorithms and heuristic optimization methods with application in volcano rapid response systems. In 2011, he joined the University of California–Berkeley as a Postdoctoral Research Scholar to work on aseismic and seismic faulting processes along the Hayward Fault. In 2013, he joined the School of Earth and Space Exploration of Arizona State University as an Assistant Professor and Head of the Radar Remote Sensing And Tectonic Geodesy Lab. In 2020, he joined Virginia Tech, where he is a remote sensing and environmental security expert with the Department of Geosciences and National Security Institute and a senior fellow of the United Nations University Institute for Water Health and Environment. Dr. Shirzaei specializes in space-borne synthetic aperture radar, groundwater resources management, green energy, and coastal hazards.

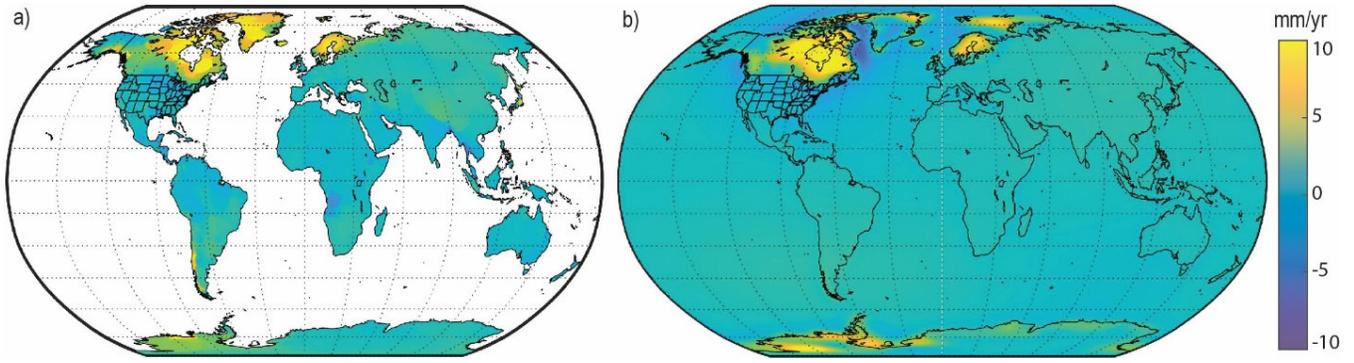

Fig. 1.
Regional models. a) Vertical land motion model derived from interpolating GNSS vertical measurements provided by [21]. b) Vertical land motion based on glacial isostatic adjustment model of [23].

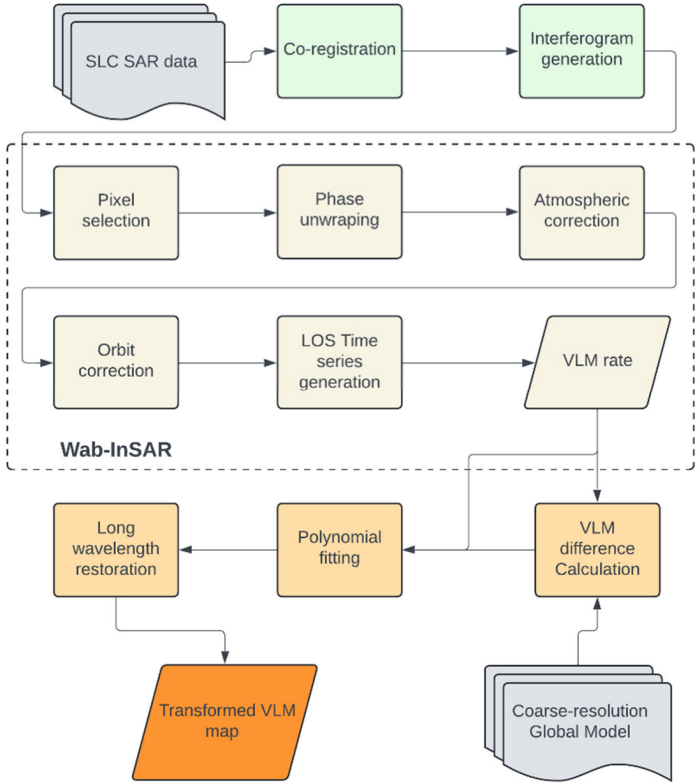

Fig. 2.
Flowchart for the transformation framework.

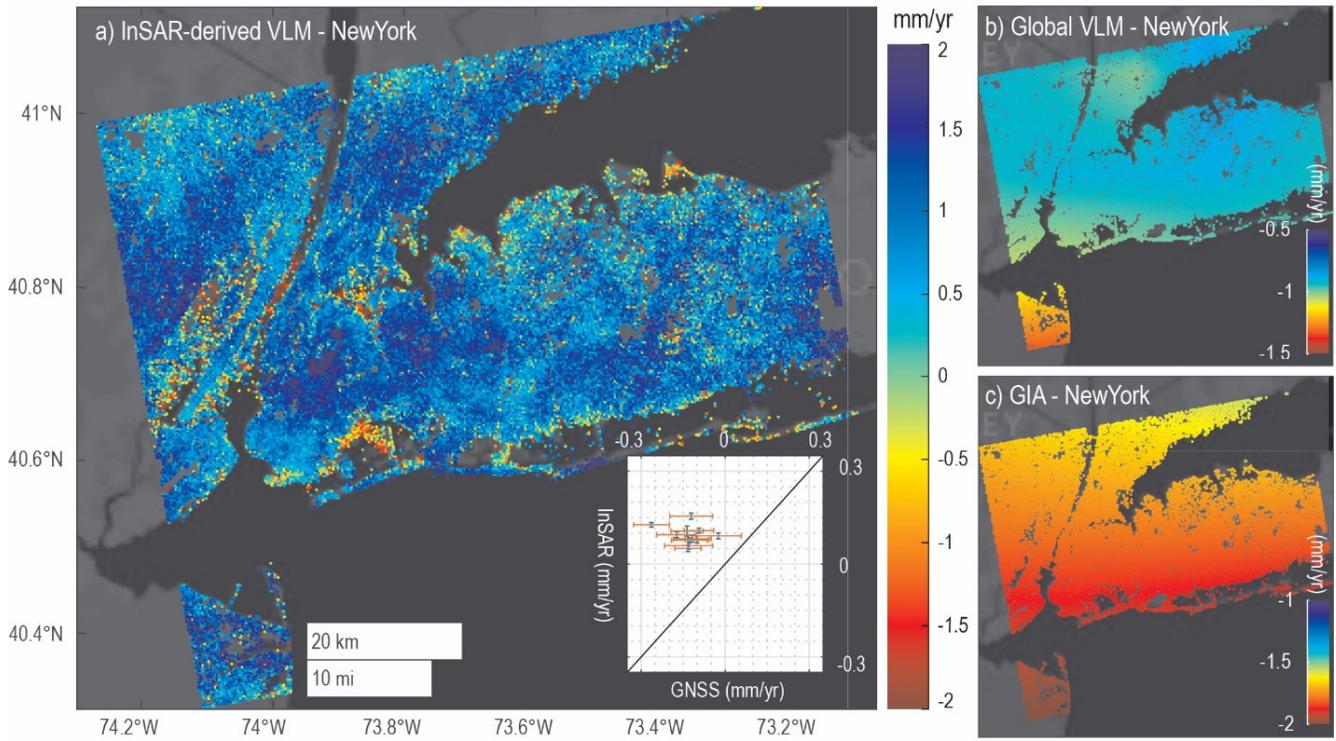

Fig. 3.
Local and global VLM models for New York City. (a) InSAR-derived VLM, referred to as the local InSAR. Inset is a bivariate plot showing the relation between the local VLM and the GNSS vertical measurements (STD for their differences = 0.07 mm/yr). (b) Global VLM model shown in Fig 1a oversampled on the location of InSAR pixels. (c) GIA model shown in Fig. 1b oversampled on the location of InSAR pixels.

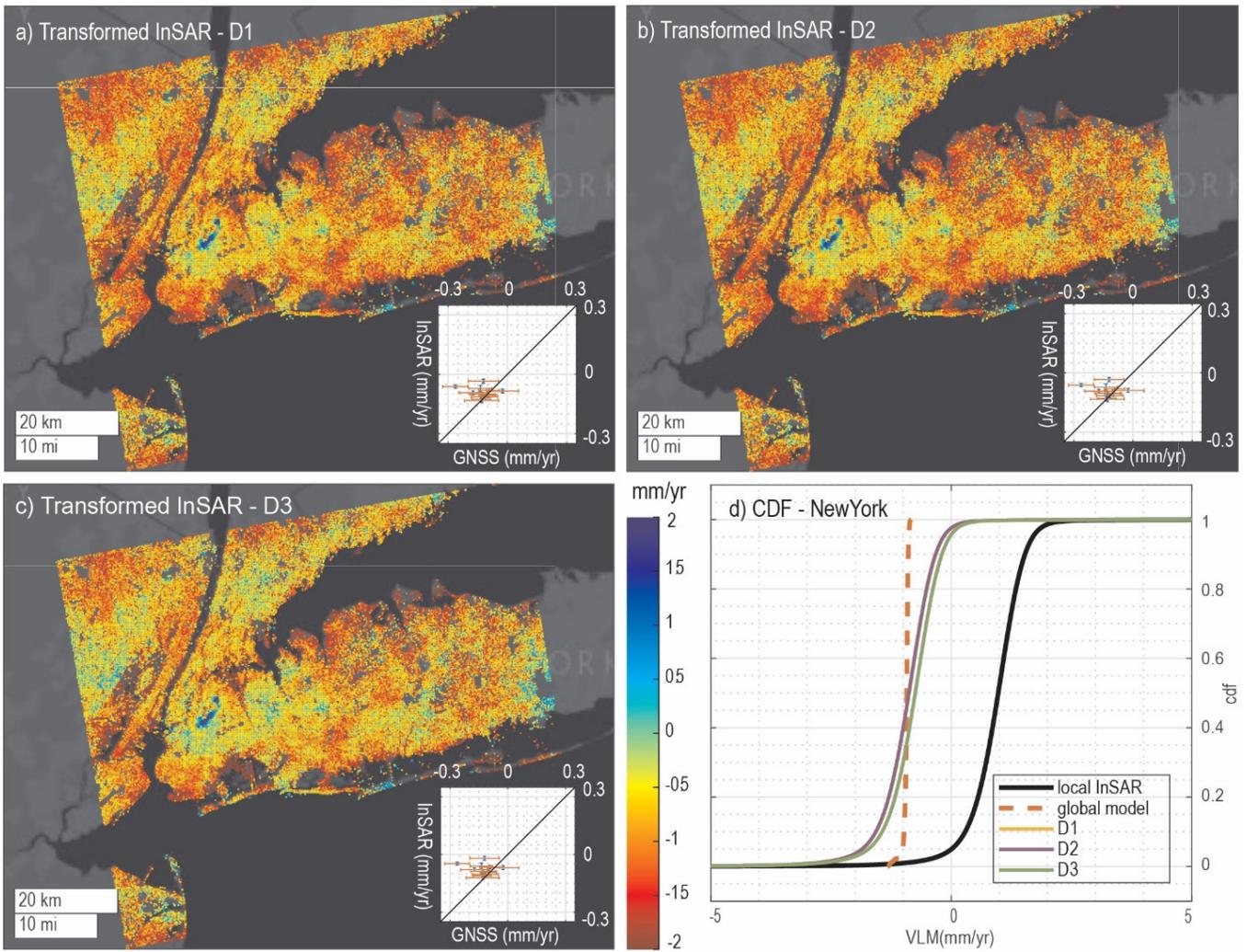

Fig. 4.
Transformed VLM for New York City and their corresponding bivariate plot showing relationships with GNSS-derived VLM. (a) Transformed VLM using a 1st-degree polynomial model (STD = 0.07 mm/yr, representing the standard deviation of differences between the modeled VLM and GNSS-derived VLM). (b) Transformed VLM using a 2nd-degree polynomial model (STD = 0.07 mm/yr, as above). (c) Transformed VLM using a 3rd-degree polynomial model (STD = 0.07 mm/yr, as above). (d) Comparison of the empirical cumulative distribution functions (CDF) for the VLM models, illustrating differences in model behavior across the transformations.

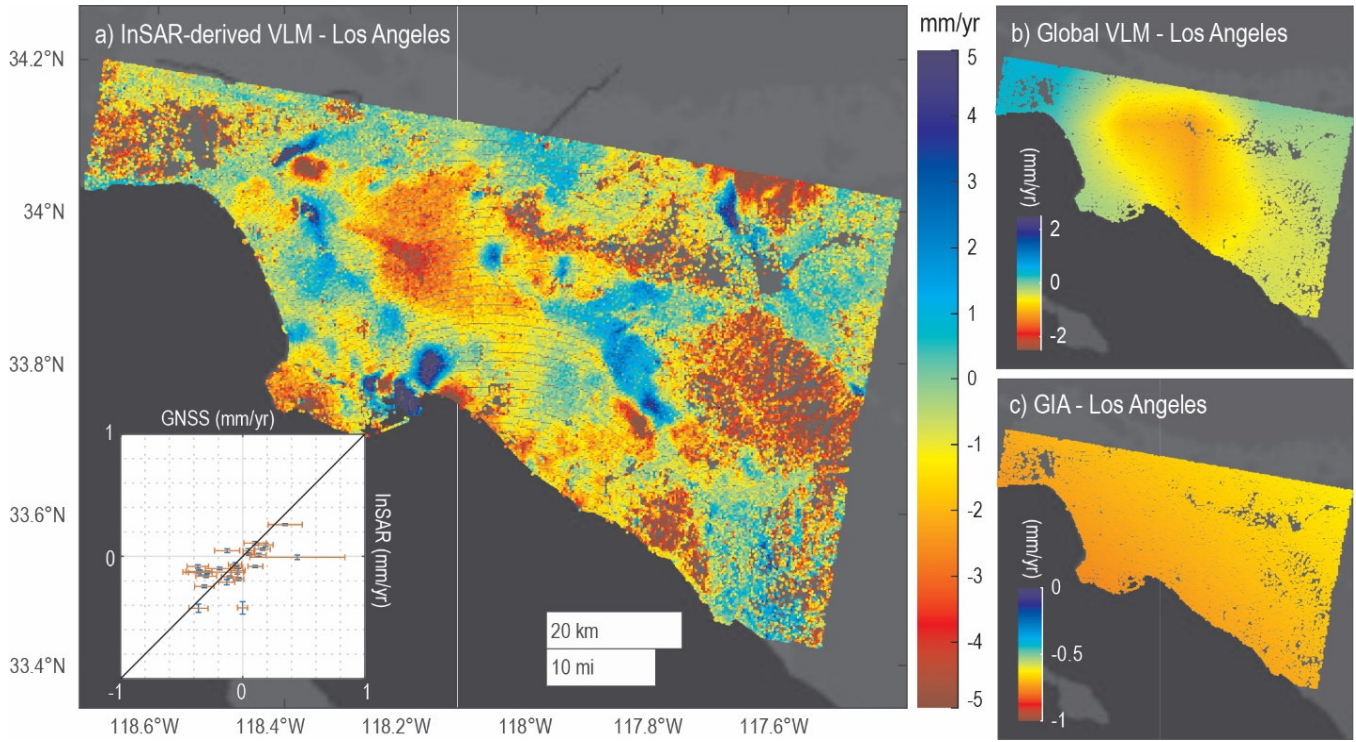

Fig. 5.
Local and global VLM models for Los Angeles. (a) InSAR-derived VLM, referred to as the local InSAR. Inset is a bivariate plot showing the relation between the local VLM and the GNSS vertical measurements (STD for their differences = 0.17 mm/yr). (b) Global VLM model shown in Fig 1a oversampled on the location of InSAR pixels. (c) GIA model shown in Fig. 1b oversampled on the location of InSAR pixels

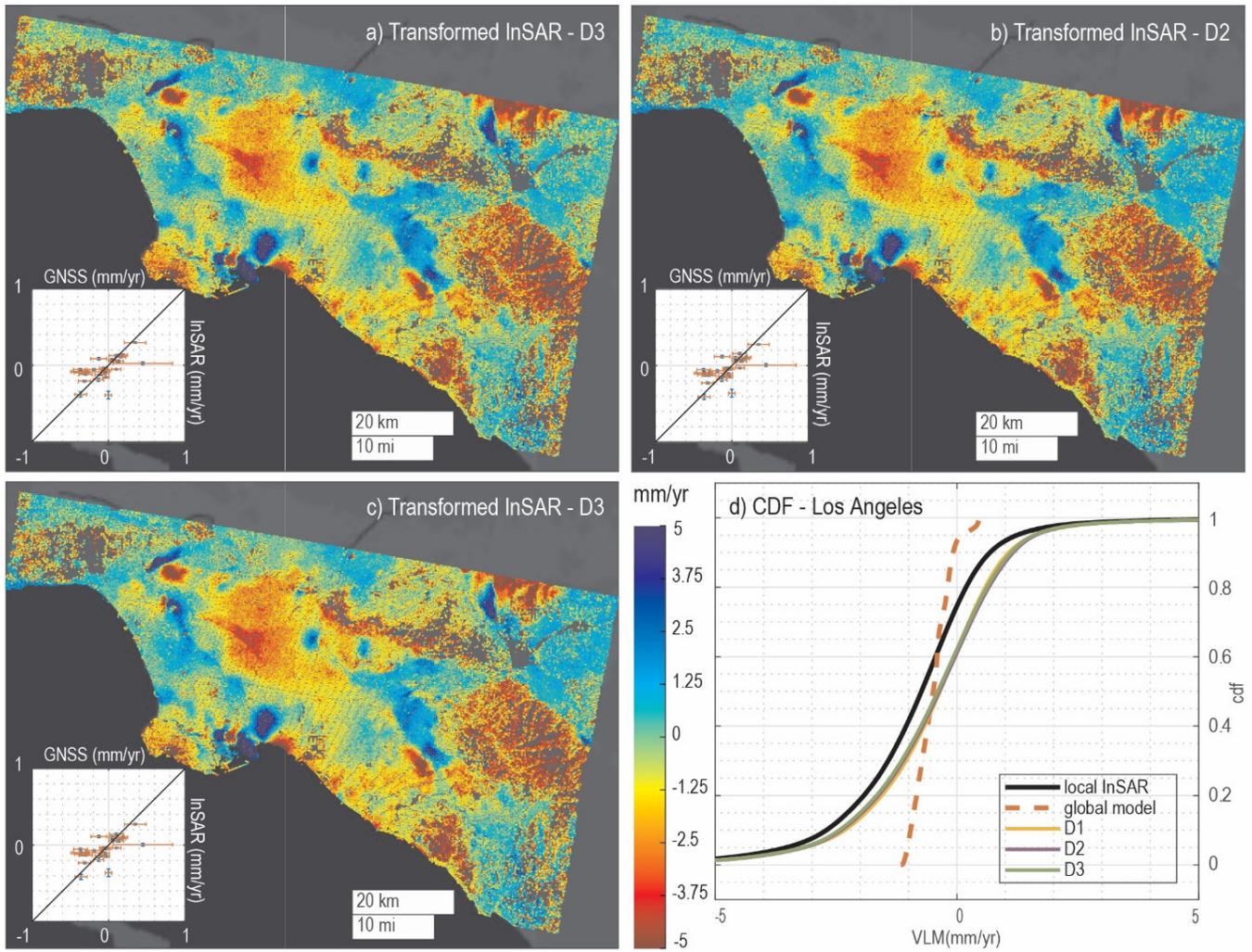

Fig. 6.
Transformed VLM for Los Angeles and their corresponding bivariate plot showing relationships with GNSS-derived VLM. (a) Transformed VLM using a 1st-degree polynomial model (STD = 0.18 mm/yr, representing the standard deviation of differences between the modeled VLM and GNSS-derived VLM). (b) Transformed VLM using a 2nd-degree polynomial model (STD = 0.17 mm/yr, as above). (c) Transformed VLM using a 3rd-degree polynomial model (STD = 0.17 mm/yr, as above). (d) Comparison of the empirical cumulative distribution functions (CDF) for the VLM models, illustrating differences in model behavior across the transformations.

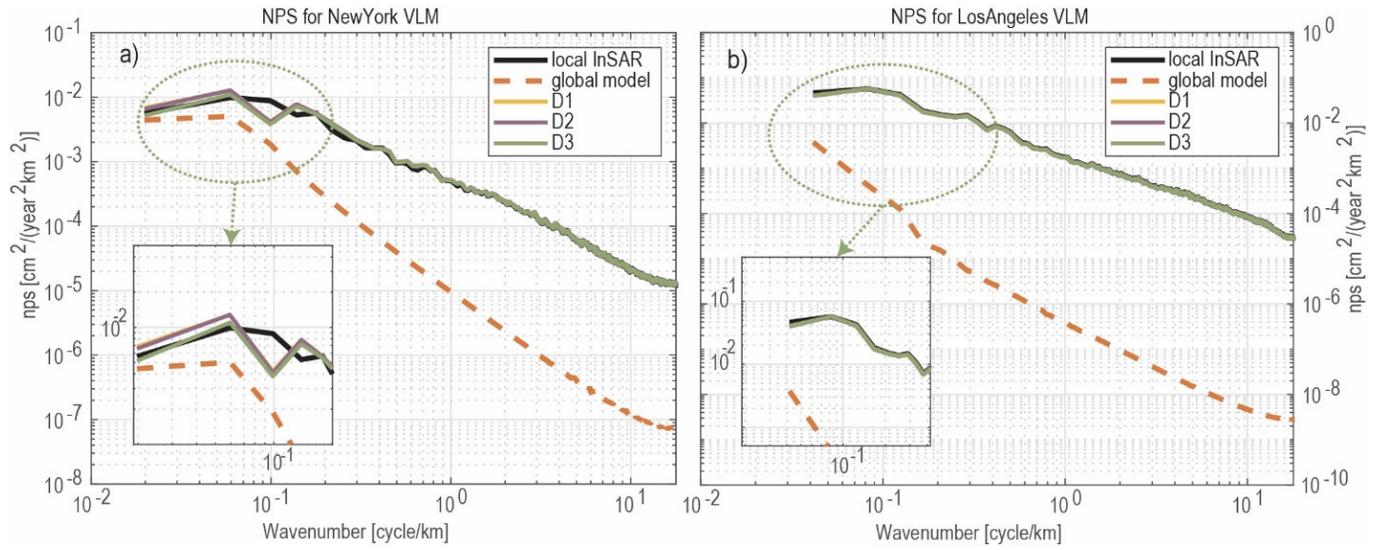

Fig. 7.
Normalized power spectrum analysis results for five VLM models. (a) New York City. (b) Los Angeles. The plots include power spectrum charts for: InSAR-derived VLM (local InSAR), global VLM model, and transformed InSAR-derived VLM using polynomial degrees 1, 2 and 3 (i.e., D1, D2, and D3).

Table 1. Model performance comparison for New York City and Los Angeles, with all models evaluated against GNSS-derived VLM. The table presents the root mean square error (RMSE), Bayesian Information Criterion (BIC), Akaike Information Criterion (AIC), mean absolute error (MAE), and standard deviation of the differences (STD). "Local InSAR" represents the InSAR-derived VLM, while D1, D2, and D3 correspond to polynomial models of degrees 1, 2, and 3, respectively. BIC and AIC are not available for the local InSAR model because it is not a parametric model fitted to the GNSS-derived VLM data.

| City | Model | RMSE (mm/yr) | BIC | AIC | MAE (mm/yr) | STD (mm/yr) |
|---|---|---|---|---|---|---|
| *New York City* | **Local InSAR** | 0.232 | -- | -- | 0.22 | 0.07 |
| | **D1** | 0.077 | -20.8 | -21.6 | 0.05 | 0.07 |
| | **D2** | 0.077 | -18.32 | -19.51 | 0.05 | 0.07 |
| | **D3** | 0.084 | -13.86 | -15.45 | 0.06 | 0.07 |
| *Los Angeles* | **Local InSAR** | 0.173 | -- | -- | 0.132 | 0.17 |
| | **D1** | 0.173 | -10.76 | -13.28 | 0.126 | 0.18 |
| | **D2** | 0.170 | -8.13 | -11.9 | 0.128 | 0.17 |
| | **D3** | 0.170 | -5.08 | -10.11 | 0.127 | 0.17 |